# Symbol Error Rates of Maximum-Likelihood Detector: Convex/Concave Behavior and Applications

Sergey Loyka, Victoria Kostina
School of Information Technology and Engineering,
University of Ottawa,
161 Louis Pasteur, Ottawa, Canada, K1N 6N5
E-mail: sergey.loyka@ieee.org

Francois Gagnon
Department of Electrical Engineering
Ecole de Technologie Superieure
1100, Notre-Dame St. West, Montreal, H3C 1K3, Canada
E-mail: francois.gagnon@etsmtl.ca

**Abstract**— Convexity/concavity properties of symbol error rates (SER) of the maximum likelihood detector operating in the AWGN channel (non-fading and fading) are studied. Generic conditions are identified under which the SER is a convex/concave function of the SNR. Universal bounds for the SER 1st and 2nd derivatives are obtained, which hold for arbitrary constellations and are tight for some of them. Applications of the results are discussed, which include optimum power allocation in spatial multiplexing systems, optimum power/time sharing to decrease or increase (jamming problem) error rate, and implication for fading channels.

## I. INTRODUCTION

Many practical problems, including optimization problems of various kinds, simplify significantly if the functions involved have some convexity/concavity properties. Not only numerical, but also analytical techniques benefit significantly if such properties hold. Powerful analytical and numerical techniques exist for convex/concave problems [1]. Significant insight into the problem is often provided by the convexity/concavity itself, even if an analytical solution is not found. Symbol error rate (SER) is an important performance measure of a digital communication systems and, as such, is often a subject to optimizations of various levels. Motivated by these arguments, this paper studies convexity/concavity properties of SER of the maximum-likelihood (ML) detector in non-fading and frequency-flat slow-fading AWGN channels. Convexity/concavity properties of ML detector error rates for binary constellations have been reported in [5]. These results are extended here to arbitrary multi-dimensional constellations. Applications of the results are discussed.

## II. SYSTEM MODEL

The standard baseband discrete-time system model with an AWGN channel, which includes matched filtering and sampling, is adopted here,

$$\mathbf{r} = \mathbf{s} + \boldsymbol{\xi} \qquad (1)$$

where $\mathbf{s}$ and $\mathbf{r}$ are $n$-dimensional vectors representing the Tx and Rx symbols respectively, $\mathbf{s} \in \{\mathbf{s}_1, \mathbf{s}_2, ..., \mathbf{s}_M\}$, a set of $M$ constellation points, $\boldsymbol{\xi}$ is the additive white Gaussian noise (AWGN), $\boldsymbol{\xi} \sim \mathcal{CN}(\mathbf{0}, \sigma_0^2 \mathbf{I})$, whose probability density function (PDF) is

$$p_\xi(\mathbf{x}) = \left(2\pi\sigma_0^2\right)^{-\frac{n}{2}} e^{-|\mathbf{x}|^2/2\sigma_0^2} \qquad (2)$$

where $\sigma_0^2$ is the noise variance per dimension, and $n$ is the constellation dimensionality; lower case bold letters denote vectors, bold capitals denote matrices, $x_i$ denotes i-th component of $\mathbf{x}$, $|\mathbf{x}|$ denotes $L_2$ norm of $\mathbf{x}$, $|\mathbf{x}| = \sqrt{\mathbf{x}^T \mathbf{x}}$, where the superscript $T$ denotes transpose, $\mathbf{x}_i$ denotes i-th vector. Frequency-flat slow-fading channels will be considered as well. The average (over the constellation points) SNR is defined as $\gamma = 1/\sigma_0^2$, which implies the appropriate normalization, $\frac{1}{M}\sum_{i=1}^{M}|\mathbf{s}_i|^2 = 1$.

Consider the maximum likelihood detector, which is equivalent to the minimum distance one in the AWGN channel, $\hat{\mathbf{s}} = \arg\min_{\mathbf{s}_i} |\mathbf{r} - \mathbf{s}_i|$. The probability of symbol error $P_{ei}$ given than $\mathbf{s} = \mathbf{s}_i$ was transmitted is $P_{ei} = \Pr[\hat{\mathbf{s}} \neq \mathbf{s}_i | \mathbf{s} = \mathbf{s}_i] = 1 - P_{ci}$, where $P_{ci}$ is the probability of correct decision. The SER averaged over all constellation points is $P_e = \sum_{i=1}^{M} P_{ei} \Pr[\mathbf{s} = \mathbf{s}_i] = 1 - P_c$. $P_{ci}$ can be expressed as

$$P_{ci} = \int_{\Omega_i} p_\xi(\mathbf{x}) d\mathbf{x} \qquad (3)$$

where $\Omega_i$ is the decision region (Voronoi region), and $\mathbf{s}_i$ corresponds to $\mathbf{x} = 0$, i.e. the origin is shifted for convenience to the constellation point $\mathbf{s}_i$. $\Omega_i$ can be expressed as a convex polyhedron [1],

$$\Omega_i = \{\mathbf{x} | \mathbf{A}\mathbf{x} \leq \mathbf{b}\}, \quad \mathbf{a}_j^T = \frac{(\mathbf{s}_j - \mathbf{s}_i)}{|\mathbf{s}_j - \mathbf{s}_i|}, \quad b_j = \frac{1}{2}|\mathbf{s}_j - \mathbf{s}_i| \qquad (4)$$

where $\mathbf{a}_j^T$ denotes j-th row of $\mathbf{A}$, and the inequality in (4) is applied component-wise.

## III. CONVEXITY OF SER IN SNR

Below we study the convexity/concavity properties of SER as a function of SNR. Only sketches of the proofs are provided here due to the page limits.

**Theorem 1**: $P_e(P_c)$ is a convex (concave) function of the SNR $\gamma$ if $n \leq 2$,

$$d^2 P_e / d\gamma^2 = P''_{e|\gamma} > 0 \leftrightarrow P''_{c|\gamma} < 0 \qquad (5)$$

Sketch of the proof is given in the Appendix.

Theorem 1 covers such popular constellations as BPSK, BFSK, QPSK, QAM, M-PSK, OOK, whose error rate convexity can also be verified directly based on known error rate expressions.

**Theorem 2**: For $n > 2$, $P_{ei}$ ($P_{ci}$) has the following convexity properties,

2.1. It is convex (concave) in the large SNR mode,

$$\gamma \geq \left(n + \sqrt{2n}\right)/d_{\min,i}^2 \qquad (6)$$

2.2. It is concave (convex) in the small SNR mode,

$$\gamma \leq \left(n - \sqrt{2n}\right)/d_{\max,i}^2 \qquad (7)$$

2.3. There are an odd number of inflection points, $P_{ci|\gamma}'' = P_{ei|\gamma}'' = 0$, in the intermediate SNR mode,

$$\left(n - \sqrt{2n}\right)/d_{\max,i}^2 \leq \gamma \leq \left(n + \sqrt{2n}\right)/d_{\min,i}^2 \qquad (8)$$

**Proof**: follows along the same lines as that of Theorem 1.

**Corollary 2.1**: Using the fact that non-negative weighted sum of convex (concave) functions is also convex (concave), the results in Theorem 2 extend directly to $P_c$ ($P_e$) by the substitutions $d_{\max,i} \to d_{\max}$ and $d_{\min,i} \to d_{\min}$, where $d_{\max} = \max_i(d_{\max,i})$ and $d_{\min} = \min_i(d_{\min,i})$, in (6)-(8).

It should be noted that the small SNR regions in (7),(8) do not exist if $d_{\max} = \infty$, i.e. unbounded $\Omega_i$.

Theorem 2 indicates that the constellation dimensionality plays an important role for concavity/convexity properties. Below we present a result which is independent of the dimensionality.

**Theorem 3**: $P_{ci}$ is log-concave in SNR for arbitrary constellation, arbitrary $n$ and any log-concave noise density (i.e. Gaussian, Laplacian, exponential, etc.)

**Proof:** via the integration theorem for log-concave functions [1, p.106].

Unfortunately, in the general case log-concavity does not extend to $P_c$ (the sum of log-concave functions is not necessarily log-concave). However, in some special cases it does.

**Corollary 3.1**: $P_c$ is log-concave under the conditions of Theorem 3 for a symmetric constellation, i.e. for $P_e = P_{e1} = P_{e2} = ... = P_{eM}$.

**Proof:** immediate from Theorem 3 since $P_c = P_{ci}$.

We note that log-concavity is a "weaker" property than concavity as the latter does not follow from the former. Yet, it is useful for many optimization problems, which can be reformulated in terms of $\log P_{ci}$.

IV. CONVEXITY OF SER IN NOISE POWER

Below we study the convexity properties of $P_{ci}$ ($P_{ei}$) as functions of the noise power, which has applications in the jamming problem.

**Theorem 4:** $P_{ei}$ has the following convexity properties in the noise power $P_N = \sigma_0^2$, for any $n$,

4.1. $P_{ei}$ is concave in the large noise mode,

$$P_N \geq d_{\max,i}^2 \left(n + 2 - \sqrt{2(n+2)}\right)^{-1} \qquad (9)$$

4.2. $P_{ei}$ is convex in the small noise mode,

$$P_N \leq d_{\min,i}^2 \left(n + 2 + \sqrt{2(n+2)}\right)^{-1} \qquad (10)$$

4.3. There are an odd number of inflection points for intermediate noise power,

$$d_{\min,i}^2 \left(n + 2 + \sqrt{2(n+2)}\right)^{-1} \leq P_N \leq d_{\max,i}^2 \left(n + 2 - \sqrt{2(n+2)}\right)^{-1} \qquad (11)$$

**Proof:** follows along the same lines as those of Theorem 1 and 2, by expressing $P_{ci}$ ($P_{ei}$) as functions of the noise power rather than the SNR.

**Corollary 4.1**: The results in Theorem 4 extend directly to $P_c$ ($P_e$) by the substitutions $d_{\max,i} \to d_{\max}$ and $d_{\min,i} \to d_{\min}$ in (9)-(11).

V. UNIVERSAL BOUNDS ON SER DERIVATIVES IN SNR

Here we explore some properties of the SER derivatives in SNR based on the results in Section III.

**Theorem 5**: The first derivative in SNR $P_{e|\gamma}'$ (and also $P_{ei|\gamma}'$) is bounded, for arbitrary constellation, as follows,

$$-\frac{c_n}{\gamma} \leq P_{e|\gamma}' \leq 0, \quad c_n = \left(\frac{n}{2}\right)^{n/2} \frac{e^{-n/2}}{\Gamma(n/2)} \qquad (12)$$

where $\Gamma(\ )$ is the gamma function.

**Proof:** follows along the lines of that of Theorem 1 and 2, by observing that the lower bound is achieved for the spherical decision region of the radius $\sqrt{n/\gamma}$ (see Corollary 5.1). The upper bound is obvious.

It should be noted that the bounds depend only on the SNR and constellation dimensionality, not on constellation geometry. They also apply to $P_{ei|\gamma}'$.

*Example*: for $n = 1$ and $n = 2$ correspondingly and arbitrary constellation geometry,

$$-\frac{1}{\sqrt{2\pi e \gamma}} \leq P_{e|\gamma}' \leq 0 \ (n=1), \quad -\frac{1}{e\gamma} \leq P_{e|\gamma}' \leq 0 \ (n=2)$$

**Corollary 5.1**: When the lower bound in (12) is applied to $P_{ei|\gamma}'$, it is achieved for the spherical decision region, $\Omega_i = C^+ = \{\mathbf{x} | \|\mathbf{x}\|^2 \leq n/\gamma\}$, of the radius $\sqrt{n/\gamma}$.

**Proof**: immediate from the proof of Theorem 5 by observing that $P_{ei|\gamma}'$ is positive outside of $C^+$.

While the spherical decision region is not often encountered in practice, it is the best possible decision region [9]. One may also expect that for decision regions of the shape close to a sphere the lower bound in (12) is tight.

**Corollary 5.2**: The asymptotic behavior of $P_{ei|\gamma}'$ and $P_{ci|\gamma}'$, which also applies to $P_{e|\gamma}'$ and $P_{c|\gamma}'$, is as follows

$$\lim_{\gamma \to \infty} P_{ei|\gamma}' = \lim_{\gamma \to \infty} P_{ci|\gamma}' = 0 \qquad (13)$$

and the convergence to the limit is uniform.

**Proof:** immediate from Theorem 5.

**Theorem 6**: The second derivative in SNR $P_{e|\gamma}''$ (and also $P_{ei|\gamma}''$) is bounded, for arbitrary constellation, as follows,

$$\frac{\beta_l}{\gamma^2} \leq P_{e|\gamma}'' \leq \frac{\beta_u}{\gamma^2} \qquad (14)$$

$$\beta_u = a_n \frac{(a_n)^{n/2} e^{-a_n}}{\Gamma(n/2)}, \quad \beta_l = -(-b_n)_+ \frac{(b_n)^{n/2} e^{-b_n}}{\Gamma(n/2)},$$

$$a_n = \tfrac{1}{2}\left(2 + \sqrt{2n}\right), \quad b_n = \tfrac{1}{2}\left(2 - \sqrt{2n}\right)$$

where $(x)_+ = x$ if $x \geq 0$ and 0 otherwise.

**Proof**: similar to that of Theorem 5, by observing that the lower and upper bounds, when applied to $P_{ei|\gamma}''$, correspond to the spherical decision regions of radii $R_l = \sqrt{(n - \sqrt{2n})_+ / \gamma}$ and $R_u = \sqrt{(n+\sqrt{2n})/\gamma}$.

*Example*: for $n = 2$ and arbitrary constellation geometry,

$$0 \leq P''_{e|\gamma} \leq \left(\frac{2}{e\gamma}\right)^2 \qquad (15)$$

**Corollary 6.1**: the lower and upper bounds in (14) are achieved for the spherical decision regions of the radii $R_l$ and $R_u$.

**Proof**: similar to that of Corollary 5.1.

**Corollary 6.2**: The asymptotic behavior of $P''_{ei|\gamma}$ and $P''_{ci|\gamma}$, which also applies to $P''_{e|\gamma}$ and $P''_{c|\gamma}$, is as follows

$$\lim_{\gamma \to \infty} P''_{ei|\gamma} = \lim_{\gamma \to \infty} P''_{ci|\gamma} = 0 \qquad (16)$$

and the convergence to the limit is uniform.

**Proof**: immediate from Theorem 6.

**Corollary 6.3**: $P_{ei}$, $P_{ci}$ (and also $P_e$, $P_c$) and their first derivatives are continuous differentiable functions of the SNR.

**Proof**: immediate from Theorems 5 and 6.

## VI. UNIVERSAL BOUNDS ON SER DERIVATIVES IN NOISE POWER

Here we explore properties of the SER derivatives in the noise power. These results parallel ones of the previous section and have similar proofs, which are omitted here for brevity.

**Theorem 7**: The first derivative in the noise power $P'_{e|P_N}$ is bounded, for arbitrary constellation, as follows,

$$0 \leq P'_{e|P_N} \leq \frac{c_n}{P_N} \qquad (17)$$

**Corollary 7.1**: The upper bound in (17) is achieved for the spherical decision region of the radius $\sqrt{nP_N}$.

**Theorem 8**: The second derivative in the noise power $P''_{e|P_N}$ is bounded, for arbitrary constellation, as follows,

$$\frac{b_l}{P_N^2} \leq P''_{e|P_N} \leq \frac{b_u}{P_N^2} \qquad (18)$$

where

$$b_u = \sqrt{\frac{n+2}{2}} \frac{(b_1)^{n/2} e^{-b_1}}{\Gamma(n/2)}, \quad b_l = -\sqrt{\frac{n+2}{2}} \frac{(b_2)^{n/2} e^{-b_2}}{\Gamma(n/2)},$$

$$b_1 = \tfrac{1}{2}\left(n+2+\sqrt{2(n+2)}\right), \quad b_2 = \tfrac{1}{2}\left(n+2-\sqrt{2(n+2)}\right)$$

**Corollary 8.1:** the lower and upper bounds in (18) are achieved for the spherical decision regions of the radii

$$R_l = \sqrt{2b_2 P_N}, \quad R_u = \sqrt{2b_1 P_N} \qquad (19)$$

with the effective SNRs $\gamma_l = R_l^2/P_N = n+2-\sqrt{2(n+2)}$ and $\gamma_u = R_u^2/P_N = n+2+\sqrt{2(n+2)}$.

**Corollary 8.2**: The asymptotic behavior of $P''_{ei|P_N}$ and $P''_{ci|P_N}$, which also applies to $P''_{e|P_N}$ and $P''_{c|P_N}$, is as follows

$$\lim_{P_N \to \infty} P''_{ei|P_N} = \lim_{P_N \to \infty} P''_{ci|P_N} = 0 \qquad (20)$$

and the convergence to the limit is uniform.

**Corollary 8.3**: $P_{ei}$, $P_{ci}$ (and also $P_e$, $P_c$) and their first derivatives are continuous differentiable functions of the noise power.

The bounds for the 1st and 2nd derivatives, both in the SNR and the noise power, can also be extended to higher-order derivatives. The analysis, however, becomes more complicated.

## VII. CONVEXITY OF AVERAGE SER IN FADING CHANNELS

Some of the convexity/concavity results above also apply to fading channels, which is explored in this section. We assume frequency-flat slow-fading channel.

**Theorem 9**: If the instantaneous SER $P_e$ is convex (concave) and the CDF of the instantaneous SNR $\gamma$ is a function of $\gamma/\gamma_0$ only,

$$CDF(\gamma) = F(\gamma/\gamma_0) \qquad (21)$$

where $\gamma_0$ is the average SNR, then the average SER $\bar{P}_e$ is convex (concave) in $\gamma_0$.

**Proof**: follows from the integral expression of the average SER, with the substitution $t = \gamma/\gamma_0$.

It should be pointed out that the convexity of $\bar{P}_e$ at the large SNR mode in a Rayleigh-fading channel can also be verified directly from the large-SNR approximation, $\bar{P}_e \approx \text{contant}/\gamma_0^k$, which is a convex function.

The equivalent to (21) condition is that the PDF of $\gamma$ can be expressed as $PDF(\gamma) = g(\gamma/\gamma_0)/\gamma_0$. The condition is not too restrictive as many popular fading channel models satisfy it, which includes Rayleigh fading channel (also with maximum-ratio combining), Rice and Nakagami fading channels. However, some channels do not satisfy (21), which includes the log-normal and composite fading channels.

## VIII. APPLICATIONS

Convexity/concavity is in high demand in any optimization problem [1]. Below we consider some of them.

*Optimum Power Allocation for the V-BLAST Algorithm*: Consider the block error rate (BLER), i.e. the probability of at least one error at the detected transmit vector, of the V-BLAST [7]:

$$P_B(\alpha_1...\alpha_m) = 1 - \prod_{i=1}^{m}(1 - P_e(\alpha_i \gamma_i)) \qquad (22)$$

where $P_e$ is the SER for the constellation in use, $\gamma_i$ is the SNR of i-th stream with uniform power allocation, $\alpha_i$ is the fraction of the total transmit power allocated to i-th stream (the uniform power allocation corresponds to $\alpha_i = 1$), $m$ is the number of streams (transmitters). Both instantaneous and average $P_e$ can be used in (22). Using the BLER as an objective, the following optimization problem can be formulated [7]:

$$\min_{\{\alpha_1...\alpha_m\}} P_B, \text{ subject to } \sum_{i=1}^{m} \alpha_i = m \qquad (23)$$

where the constraint insures that the total transmit power is fixed.

**Theorem 10:** The optimization problem in (23) has a unique solution for either: (i) 1-D or 2-D constellations in terms of instantaneous or average (in Rayleigh, Rice and Nakagami-fading channels) BLER, or (ii) for *M*-D symmetric constellations, $M \geq 1$, in terms of instantaneous BLER.

**Proof:** note that the problem in (23) is equivalent to $\max_{\{\alpha_1...\alpha_m\}} \sum_{i=1}^{m} \ln(1-P_e(\alpha_i \gamma_i))$. If $P_e$ is convex, $(1-P_e)$ and $\ln(1-P_e)$ are concave [1]. Thus, the objective function is concave and hence the problem has a unique solution. By Theorem 1 and 9, this holds for 1-D or 2-D constellations in the AWGN channel, or Rayleigh, Rice, or Nakagami fading channels if the average BLER is used. For $M \geq 1$ and a symmetric constellation, the uniqueness in terms of instantaneous BLER follows from Corollary 3.1.

*Optimum Power/Time Sharing for a Jammer*: This section extends the corresponding results in [5] to non-binary multi-dimensional constellations. Since the proofs of these results follow along the same lines as those in [5], we omit them for brevity.

Considering $P_e$ as a function of $P_N$, one may formulate the following jamming optimization problem using power/time sharing:

$$\max_{n, \{\alpha_1...\alpha_n\}, \{P_{N1}...P_{Nn}\}} \sum_{i=1}^{n} \alpha_i P_e(P_{Ni})$$
$$\text{subject to } \sum_{i=1}^{m} \alpha_i = 1, \ \sum_{i=1}^{m} \alpha_i P_{Ni} = P_N \quad (24)$$

where the jammer splits its transmission into $n$ sub-intervals, $\alpha_i$ being the fractional length of i-th sub-interval and $P_{Ni}$ is its noise (jammer) power, with the total noise power $= P_N$, and n is the number of sub-intervals. The objective function in (24) is the SER over the whole transmission interval. An immediate conclusion from (24) is that if $P_e(P_N)$ is concave, the power/time sharing does not help, i.e. the best strategy is no sharing: $n=1$, $\alpha_1 = 1$, $P_{N1} = P_N$. This can be seen from the basic concavity inequality,

$$\sum_{i=1}^{n} \alpha_i P_e(P_{Ni}) \leq P_e\left(\sum_{i=1}^{n} \alpha_i P_{Ni}\right) = P_e(P_N) \quad (25)$$

Theorem 4 ensures that the optimization is possible. The optimum $n$ follows immediately from Caratheodory theorem [5, 6]: $n \leq 2$, where $n=1$ corresponds to no sharing so that the only non-trivial solution is $n=2$, i.e. two power levels are enough to achieve the optimum. Let us denote the maximum in (24) as $\tilde{P}_e(P_N)$, where "~" denotes optimality. Similarly to [5], it has simple characterization:

**Lemma 1:** $\tilde{P}_e(P_N)$ is concave.

**Proof**: by contradiction[1]. If it is not concave, one can apply the sharing in (24) again to increase it. But that is impossible as two consecutive sharings are equivalent to a single one and hence the second one does not help. Thus, $\tilde{P}_e(P_N)$ has to be concave, in which case second sharing does not help, as (25) indicates. Q.E.D.

It also follows that $\tilde{P}_e(P_N)$ is the smallest concave function that upper-bounds $P_e(P_N)$ [1,5,6]. This fact, however, is immaterial for our problem as we try to maximize $P_e$ so larger functions are naturally welcome.

Before finding the optimal solution, we give a sub-optimal one, which is however simpler to characterize. For clarity of exposition, we consider here the case of a single inflection point ($P_0$) only; the extension to the general case is

---

[1] The original proof in [5] relied on an elaborate argument. The contradiction-type proof given here is much simpler.

straightforward. If follows from Theorem 4 that,

$$P''_{e|P_N} > 0, \ P_N < P_0$$
$$P''_{e|P_N} < 0, \ P_N > P_0 \quad (26)$$

and the sub-optimum sharing is as follows:

**Theorem 11**: The sub-optimum solution to (24) is to use the single power level (always "on") $P_{N1} = P_N$ if $P_N \geq P_0$, and "on-off" strategy with the on-interval $\alpha_1 = P_N/P_0$, $P_{N1} = P_0$ if $P_N < P_0$,

$$\begin{cases} \alpha_i, P_{Ni}, \\ i=1...n \end{cases} = \begin{cases} n=1, \ \alpha_1 = 1, \ P_{N1} = P_N & \text{if } P_N \geq P_0 \\ n=2, \ \alpha_1 = \dfrac{P_N}{P_0}, \ P_{N1} = P_0, \ P_{N2} = 0 & \text{if } P_N < P_0 \end{cases} \quad (27)$$

which achieves the following SER,

$$\tilde{P}_e(P_N) = \begin{cases} P_e(P_N), & P_N \geq P_0 \\ P_e(P_0)P_N/P_0, & P_N < P_0 \end{cases} \quad (28)$$

**Proof**: it is straightforward to verify that (28) corresponds to the strategy in (27). Using (26), it follows that $\tilde{P}_e(P_N) \geq P_e(P_N)$. Thus, (27) is indeed a better strategy than no sharing. Q.E.D.

Intuitive explanation for (27) is that one eliminates the convex part of $P_e(P_N)$ by time/power sharing and the concave part is left intact (no optimization is required there). Indeed, it can be verified that $\tilde{P}''_{e|P_N} = 0$ if $P_N < P_0$ and $\tilde{P}''_{e|P_N} < 0$ if $P_N > P_0$. The solution in (27) is not optimum since the first derivative of $\tilde{P}_e(P_N)$ is discontinuous at $P_N = P_0$ and $\tilde{P}''_{e|P_N}(P_0) = +\infty$ (unless $P'_{e|P_N}(P_0) = P_e(P_0)/P_0$, in which case (27) gives the optimum solution) so that $\tilde{P}_e(P_N)$ is not concave, which means that further optimization is possible.

It follows that the optimal solution for the single inflection point case is the same as that in [5, Theorem 3] (because it is based only on the convexity/concavity properties of the problem, which were demonstrated above), and it is identical to (27) with a differently-defined threshold $P_0$.

*Optimum Time/Power Sharing for the Transmitter*: Similarly to the jammer problem above, the optimization problem can be formulated for the transmitter, with the objective to reduce the SER. In fact, these two problems are equivalent, via the substitutions,

$$P_c \rightarrow P_e, \ \gamma \rightarrow P_N \quad (29)$$

For completeness, we formulate below the main results.

**Theorem 12**: If $P_c(\gamma)$ is concave, e.g. for 1-D and 2-D constellations, the optimum transmission strategy is always "on", without sharing (i.e. power/time sharing does not help to reduce the SER, which was the case in [5] for a binary modulation). If $P_c(\gamma)$ is not concave, e.g. for some M-D constellations, $M \geq 3$, (i) the sub-optimum transmitter strategy is given by Theorem 11, and (ii) the optimum transmitter strategy is given by [5, Theorem 3], both with the substitutions in (29).

Comparing these results to those in the previous section, we conclude that the jammer is in better position compared to the transmitter for 1-D and 2-D constellations.

*Implication for Fading Channels:* The following result is a straightforward consequence of the basic convexity inequality and the results in Section III.

**Theorem 13**: If $P_e(\gamma)$ is convex in the non-fading AWGN channel, e.g for 1-D and 2-D constellations, fading never reduces the SER ("fading is never good"),

$$\overline{P_e(\gamma)} \geq P_e(\bar{\gamma}) \qquad (30)$$

where $\bar{x}$ denotes mean value of $x$.

\*\*\*

It was recently brought to our attention that the convexity/concavity properties of error rates have also implications for the inter-symbol interference problem [8].

## X. APPENDIX

**Proof of Theorem 1 (sketch):** consider $P''_{ci|\gamma}$, which can be expressed as

$$P''_{ci|\gamma} = \int_{\Omega_i} \frac{d^2 p_\xi(\mathbf{x})}{d\gamma^2} d\mathbf{x} \qquad (31)$$

where the derivative is

$$\frac{d^2 p_\xi(\mathbf{x})}{d\gamma^2} = \frac{1}{4}\left(\frac{\gamma}{2\pi}\right)^{n/2} e^{-\gamma|\mathbf{x}|^2/2} f(|\mathbf{x}|^2) \qquad (32)$$

and $f(t) = (t - \alpha_1/\gamma)(t - \alpha_2/\gamma)$, $\alpha_1 = n + \sqrt{2n} > 0$, $\alpha_2 = n - \sqrt{2n} \leq 0$, so that $f(|\mathbf{x}|^2) \leq 0$ if $|\mathbf{x}|^2 \leq \alpha_1/\gamma$, and $f(|\mathbf{x}|^2) > 0$ otherwise. Consider three different cases.

(i) If $d^2_{\max,i} \leq \alpha_1/\gamma$, where $d_{\max,i}$ is the maximum distance from the origin to the boundary of $\Omega_i$, then $f(|\mathbf{x}|^2) \leq 0$ $\forall \mathbf{x} \in \Omega_i$ so that the integral in (31) is clearly negative and (5) follows. Fig. 1 illustrates this case. This is a low-SNR mode since $\gamma \leq \alpha_1/d^2_{\max,i}$.

(ii) If $d^2_{\min,i} \geq \alpha_1/\gamma$, where $d_{\min,i} = \min_j(b_j)$ is the minimum distance from the origin to the boundary of $\Omega_i$, then $f(|\mathbf{x}|^2) \geq 0$ $\forall \mathbf{x} \in (\mathbf{R}^n - \Omega_i)$, where $\mathbf{R}^n$ is the $n$-dimensional space, and $(\mathbf{R}^n - \Omega_i) = \{\mathbf{x} | \mathbf{x} \notin \Omega_i\}$ is the complement of $\Omega_i$. The integral in (31) can be upper bounded as

$$P''_{ci|\gamma} = \int_{\Omega_i} \frac{d^2 p_\xi(\mathbf{x})}{d\gamma^2} d\mathbf{x} < \int_{\mathbf{R}^n} \frac{d^2 p_\xi(\mathbf{x})}{d\gamma^2} d\mathbf{x} = 0 \qquad (33)$$

where we have used the fact that $\int_{\mathbf{R}^n} p_\xi(\mathbf{x}) d\mathbf{x} = 1$. This is a large-SNR mode since $\gamma \geq \alpha_1/d^2_{\min,i}$.

(iii) The last case of $d^2_{\min,i} < \alpha_1/\gamma < d^2_{\max,i}$ is illustrated in Fig. 2. Separate the decision region $\Omega_i$ into two sub-regions, $\Omega_i = \Omega_a + \Omega_b$, $\Omega_a = \Omega_i - \Omega_i \cap \Omega_{con}$, $\Omega_b = \Omega_i \cap \Omega_{con}$, where $\Omega_{con}$ is (are) the cone(s) whose base(s) is (are) the intersection(s) of the planes $\mathbf{a}_j^T \mathbf{x} = b_j$ and the ball $|\mathbf{x}|^2 \leq \alpha_1/\gamma$; the vertex of the cone(s) is the origin $\mathbf{x} = 0$. Clearly, the integral over $\Omega_b$ is negative. The integral over $\Omega_a$ can be bounded as

$$\int_{\Omega_a} \frac{d^2 p_\xi(\mathbf{x})}{d\gamma^2} d\mathbf{x} < \int_{(\mathbf{R}^n - \Omega_{con})} \frac{d^2 p_\xi(\mathbf{x})}{d\gamma^2} d\mathbf{x} = 0 \qquad (34)$$

where the inequality follows from the fact that $f(|\mathbf{x}|^2) > 0$ $\forall \mathbf{x} \in (\mathbf{R}^n - \Omega_{con} \cup \Omega_i)$, and the equality follows from the fact that the integral over $(\mathbf{R}^n - \Omega_{con})$ is independent of SNR. Combining the bounds for the integrals over $\Omega_a$ and $\Omega_b$, one obtains (5).

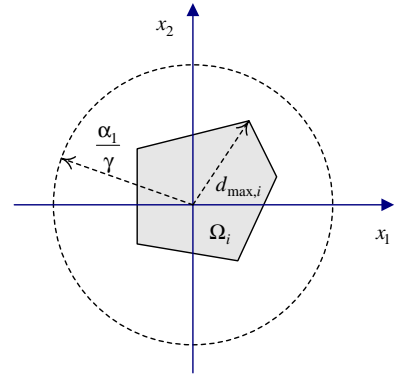

**Fig. 1. Two-dimensional illustration of the problem geometry for Case (i). The decision region $\Omega_i$ is shaded.**

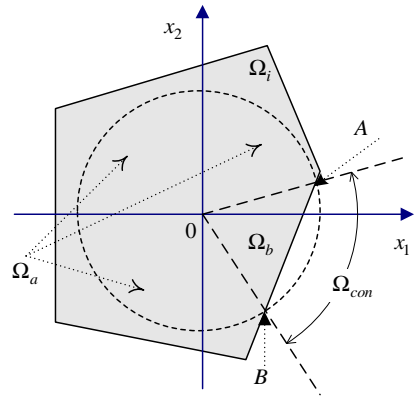

**Fig. 2. Two-dimentional illustration of the problem geometry for Case (iii). The cone $\Omega_{con}$ is build on the OA and OB rays. $\Omega_b$ is the triangle AOB.**